# Nucleation of a stable solid from melt in the presence of multiple metastable intermediate phases: Wetting, Ostwald step rule and vanishing polymorphs

**Mantu Santra, Rakesh S. Singh and Biman Bagchi***

**Solid State & Structural Chemistry Unit**
**Indian Institute of Science, Bangalore 560012, India.**

*Email: bbagchi@sscu.iisc.ernet.in

## Abstract

In many systems, nucleation of a stable solid may occur in the presence of other (often more than one) metastable phases. These may be polymorphic solids or even liquid phases. Sometimes the metastable phase might have lower free energy minimum than the liquid but higher than the stable solid phase minimum and have characteristics in between the parent liquid and the globally stable solid phase. In such cases, nucleation of the solid phase from the melt may be facilitated by the metastable phase because the latter can "wet" the interface between the parent and the daughter phases, even though there may be no signature of the existence of metastable phase in the thermodynamic properties of the parent liquid and the stable solid phase. Straightforward application of classical nucleation theory (CNT) is flawed here as it overestimates the nucleation barrier since surface tension is overestimated (by neglecting the metastable phases of intermediate order) while the thermodynamic free energy gap between daughter and parent phases remains unchanged. In this work we discuss a density functional theory (DFT) based statistical mechanical approach to explore and quantify such facilitation. We construct a simple order parameter dependent free energy surface that we then use in DFT to calculate (i) the order parameter profile, (ii) the overall nucleation free energy barrier and (iii) the surface tension between the parent liquid and the metastable solid and also parent liquid and stable solid phases. The theory indeed finds that the nucleation free energy barrier can decrease significantly in the presence of wetting. This approach can provide a microscopic explanation of Ostwald step rule and the well-known phenomenon of "disappearing polymorphs" that depends on temperature and other thermodynamic conditions. Theory reveals a diverse scenario for phase transformation kinetics some of which may be explored via modern nanoscopic synthetic methods.

## I. Introduction

In the random first order transition (RFOT) theory of glass transition **[1]**, nucleation of a liquid droplet within a glass/amorphous phase was proposed as the basic relaxation mechanism. The authors employed an unusual size dependence of the surface tension in the form of $\gamma(R) = \gamma_0 / R^{1/2}$, where $\gamma(R)$ is the size (R) dependent surface tension of the droplet-glass interface. While curvature dependence of surface tension is often derived in terms of Tolman's length, this dependence has a completely different origin. This square root dependence comes from the idea of random Ising model where nucleus of new phase can be wetted by many phases of intermediate (between daughter and parent phases) order.

When this unusual square root dependence of surface tension is substituted back in classical nucleation theory (CNT), it gives rise to well known Adam-Gibbs (AG) **[2]** relation between relaxation time and configuration entropy ($s_c$), thus providing a simple and elegant derivation of this famous relationship. The original derivation of AG and many subsequent studies have focused on the cooperatively rearranging regions (CRR) that forms the basis of AG relation. These CRRs are often identified with a correlation length in a deeply supercooled liquid. In the AG picture the size of the CRR increases rapidly as the liquid approaches the glass transition and configurational entropy approaches zero. In the Xia-Wolynes treatment **[1]**, the size of the critical nucleus grows as $s_c^{-2/3}$ as the configuration entropy ($s_c$) decreases. However, the derivation of Xia and Wolynes [1] apparently does not require such a growing correlation length, this picture is quite different and based on the ideas of nucleation and first order phase transitions.

However, the peculiar size dependence of the surface tension has not been investigated in detail. As mentioned in the one of the preceding paragraphs, the peculiar size dependence arises by invoking the concept of "wetting" of the interface between the growing liquid and the parent glass. The concept of wetting is a well known phenomenon and play direct or indirect role in many theoretical studies and has been often observed experimentally and in computer simulation studies [3-8]. For example, a face centered cubic (fcc) solid phase may form where body centered cubic (bcc) solid remains a metastable solid phase.

In this paper, we shall take the cue from Xia and Wolynes [1] but employed density functional theory (DFT) to study the effect of intermediate phase on the nucleation and growth scenario in complex systems. We demonstrate that wetting of interface by phases of intermediate order can dramatically reduce the value of surface tension and under certain conditions the surface tension decreases inversely with number of phases.

In an elegant application of irreversible thermodynamics van Santen [9] showed that the formation of stable phase is facilitated by the presence of intermediate phases. This is intimately connected with Ostwald step rule (OSR) [10, 11]. This quantifies the condition under which different phases will appear.

Ostwald argued that the formation of a phase is not determined by its absolute stability but by closeness of the growing phase to the parent phase [10]. Although Ostwald did not mention explicitly but what he meant by "closeness" is that the surface tension determines the growth. A demonstration of this fact comes from critical phenomena where surface tension varies as $\gamma \sim (\Delta\rho)^4$.

Thus we see two very different problems in the area of scientific research (namely glass transition and the synthesis of solids) can be related very intimately. The formation of a new solid is often wetted by metastable solids to lower surface tension. The intermediate phase can be metastable with respect to either both the solid phases or only the final solid (in the case of crystallization) and liquid (in the case of glass transition). Additionally, the number of phases involved in wetting can have dramatic effects. In a highly interesting study Granasy and Oxtoby [12] already studied this scenario in the presence of one metastable phase. Although they did not mention the Ostwald step rule explicitly (surprising omission), their analysis essentially provides a beautiful elucidation of condition underlying validity of Ostwald step rule (OSR). However, Granasy and Oxtoby [12] considered one metastable intermediate phase that distinguishes their work from the work of Xia and Wolynes [1] who had multiple phases wetting the interface.

Intervention by multiple intermediate phases lowers the surface tension, the unusual $1/\sqrt{R}$ is not clear yet. Such a relation can be rigorously valid only asymptotically and can require very broad interface such that many intermediate phases can be accommodated. Granasy and Oxtoby [12] did observe the conditions where the interface becomes very broad and it is also observed that surface decreases with the width of the interface as $\gamma \sim 1/w^2$ [13] (without wetting), one would naturally suspect that Xia-Wolynes expression could be valid under certain conditions which need to be explored in details.

Furthermore, protein folding and crystallization from melt to solids with multiple polymorphs – the two phenomena of current interest in biology and materials science share a common physical chemistry basis. Both have similar rugged energy landscape arising from definite entropy-enthalpy relationships **[14-16]**. Energy landscape paradigm of condensed matter science **[15, 17, 18]** assumes existence of multiple free energy minima in the configurational space, with the minima (in principle) being multiply connected. Transition between such minima is described in terms of free energy barriers along a chosen path. We discuss here that choice of the efficient path can be dictated by thermodynamic conditions. Therefore, thermodynamics (free energy minima) and kinetics (barriers) are intimately correlated in the energy landscape.

In the case of crystallization of a solid from liquid or melt, we need order parameters that uniquely identify different structures. That is, ideally each minimum should be characterized by a set of values of the order parameters and should correspond to a unique structure. In the Ramakrishnan-Yusouff density functional theory of freezing **[19]**, the order parameters are the density components evaluated at the reciprocal lattice vectors of the solid, along with the fractional density change. It is relatively easy to make either equilibrium or a dynamic calculation of freezing to different lattice types. That is, one can explain why argon freezes into an fcc lattice while liquid sodium freezes into a bcc solid. However, the situation is far more difficult in the case of complex solids like zeolites **[20]**. Here we do not have the information about the liquid structure necessary for a microscopic theory.

An elegant illustrative example of Ostwald step rule is recently provided in the formation of the thermodynamically stable crystalline form of LiFePO$_4$ (olivine structure) via multiple metastable intermediate crystalline phases **[16]**. In the energy landscape picture the surface tension and difference between the minimum of free energy corresponding to two structures play an important role in determining nucleation barrier between two metastable or one metastable and one stable solid phases. However, it is very hard to obtain surface tension between two metastable solid phases.

In the energy landscape view, the polymorphs are the inherent structures of the sol phase and should be obtained when the vibrational degrees of freedom and the kinetic energy are removed from the molecules. Thus, the polymorphs form a rugged landscape with the most stable structure at the bottom of energy ladder – just like the rugged folding funnel of a protein developed by Onuchic, Wolynes and coworkers **[15]**.

When we consider formation of the metastable solid MS (which we assume to be the closest to the sol phase), then the free energy gap $\varDelta G_V$, is lower than the most stable phase referred as stable solid (SS). Thus, according to CNT **[21-26]**, the only way the phase MS can precipitate out at any temperature is to have such a lower surface tension that the nucleation barrier is lowest for MS. Because of lower solid-liquid surface tension leading to lower nucleation barrier, metastable solids are kinetically favored. Thus, kinetics seems to play a very dominant role. We must note that even in simple systems the validity of CNT (at least at high supersaturation) itself is questionable **[27-32]**.

At high temperature, the following proposed scenario holds. Since the energy of the system is high, it can probe all the minima of the system. Even if it gets trapped in a low lying minimum, like in M1 or M2 phase, it can escape from the minimum before the phase grows to macroscopic size. In other words, when nucleus forms, it can melt within a time comparable to the relaxation time of the system. It of course gets trapped many times in the low lying minima, and gets out again and again. When it gets trapped in the deep minimum of the most stable phase it can grow. However, when the temperature is low, it gets trapped in the closest minimum as envisaged by Ostwald.

In the next section we construct a density functional theory, which provides a quantitative explanation of the sequential formation of metastable states before transforming to the most stable phase.

## II. One metastable intermediate phase: Density functional theory of surface tension and nucleation barrier

The classical density functional theory was introduced by Lebowitz and Percus **[33]** and Stillinger and Buff **[34]** and has been advanced by many other in last five decades **[35-38]**. Oxtoby and coworkers applied extensively the density functional theory to study the nucleation processes in many simple and complex systems **[36-38]**. In this work first we shall discuss the one order parameter description to bring out the generality of the problem then a two order parameter description is provided. In the next section we shall describe density functional theory of surface tension and nucleation barrier in presence of one metastable intermediate phase. A schematic representation of free energy surface is provided in **Fig. 1**.

### A. Numerical implementation of the effects of wetting within a one order parameter theory

The ideas articulated in the previous section can be nicely verified within a one order parameter theory, using density functional theory. The proposed free energy functional for three phases, low density melt, intermediate metastable solid (MS) phase and high density stable solid (SS) phase are

$$\Omega_i[\rho(\mathbf{r})] = \int d\mathbf{r}\left[f_i(\rho(\mathbf{r})) - \mu\rho(\mathbf{r})\right] + \frac{1}{2}\int d\mathbf{r}\left[K_\rho(\nabla\rho(\mathbf{r}))^2\right] \quad (1)$$

where $f_i$ is local Helmholtz free energy density function of the average number density $\rho(\mathbf{r})$ of the *ith* phase and $\mu$ is the chemical potential. Here '$i$' indicates respective phases as $i$ = M stands for melt, $i$ = MS for intermediate solid and $i$ = SS for stable solid phase. The last term (square gradient term) accounts for the nonlocal effects in the system due to inhomogeneity in density order parameter. $K_\rho$ is related to density correlation length. The Helmholtz free energy density for each phase is

$$f_i(\rho) = k_B T a_i (\rho - \rho_i)^2 + f_{i,0} \quad (2)$$

where $i = M$ for melt, $MS$ for metastable solid and $SS$ for the stable solid phase respectively. Here $k_B$ is the Boltzmann's constant, $T$ is the absolute temperature. The value of the parameters are, $a_M$ = 1500, $a_{MS}$ = 2000, $a_{SS}$ = 2500, equilibrium densities are $\rho_M$ = 0.88, $\rho_{MS}$ = 0.97, $\rho_{SS}$ = 1.05, $f_{M,0}$ = 0.0, $f_{SS,0}$ = 0.80, and $f_{MS,0}$ is varied from 0.6 to 20.0.

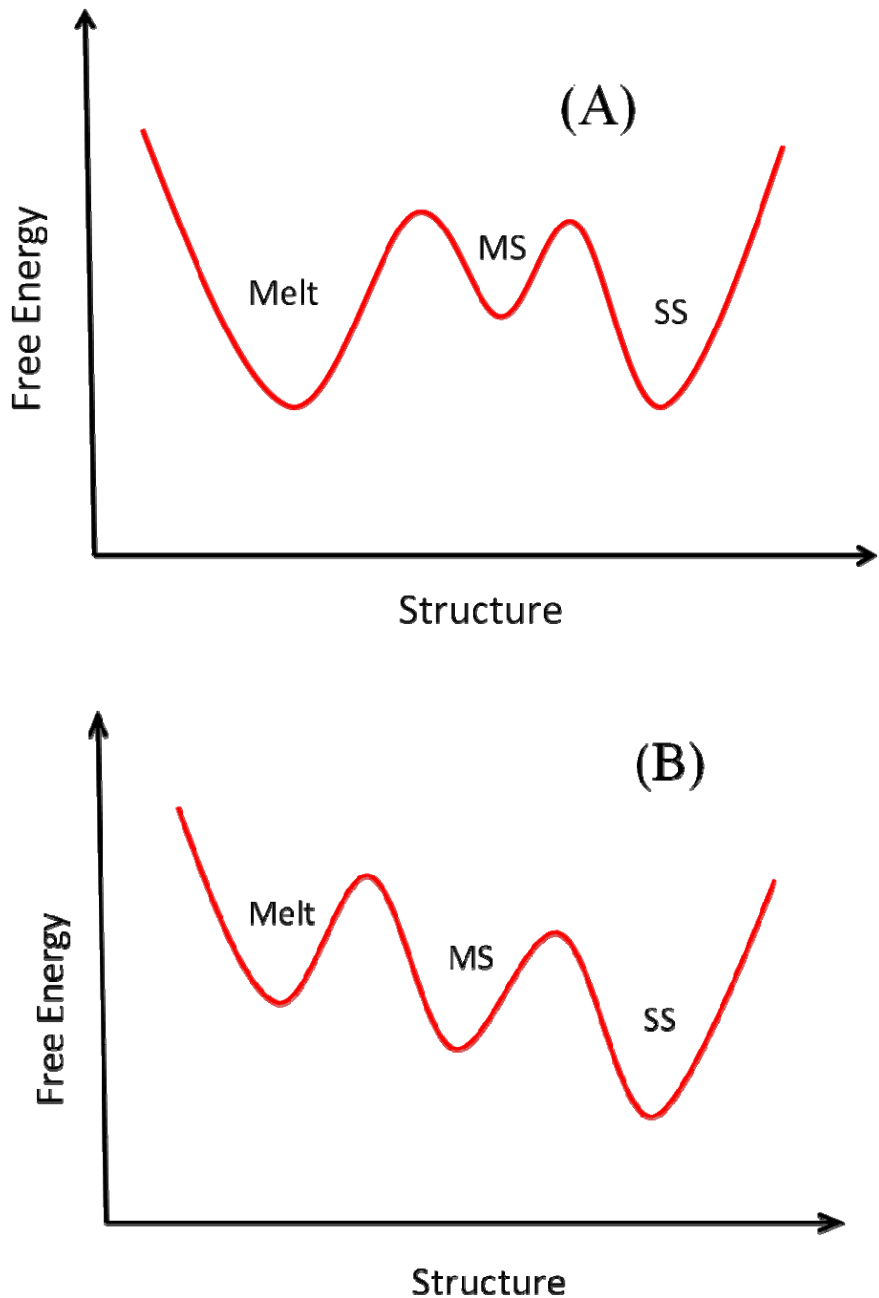

**Figure – 1.** Schematic free energy landscape of crystallization of stable solid (SS) from melt in the presence of a metatsable intermediate solid (MS) phase. (A) The case when the melt and stable solid phases are at coexistence and the intermediate phase is metastable with respect to both (melt and stable solid). (B) The case when intermediate phae is stable with respect to melt but metastable with respect to the globally stable solid phase.

## B. Surface tension

In order to get the surface tension between coexisting phases first we need to get the equilibrium densities of coexisting phases. This can be determined by equating chemical potential and thermodynamic grand potential density (pressure) of two phases.

$$\mu_\alpha(\rho_\alpha) = \mu_\beta(\rho_\beta) \text{ and } \omega_\alpha(\rho_\alpha) = \omega_\beta(\rho_\beta) \tag{3}$$

where $\mu_i = \left(\dfrac{\partial f_i(\rho)}{\partial \rho}\right)_T$ and $\omega_i = f_i - \mu_i \rho_i$.

The above two conditions ensure that the system is in both thermodynamic and mechanical equilibrium.

We can evaluate the values of the surface tension between the coexisting phases for a planar interface along z-axis by solving the Euler-Lagrange equations associated with following equilibrium conditions

$$\frac{\delta\Omega[\rho(z)]}{\delta\rho(z)} = 0, \tag{4}$$

where $\Omega[\rho(z)]$ is the grand canonical free energy functional corresponding to the inhomogeneous system with density profile $\rho(z)$,

$$\Omega[\rho(z)] = \int dz \left[f(\rho(z)) - \mu\rho(z)\right] + \frac{1}{2}\int dz \left[K_\rho (\nabla\rho(z))^2\right], \tag{5}$$

where $f = \min\{f_i\}$.

The density profiles shown in **Fig. 2A** are obtained by solving the corresponding Euler-Lagrange equation (**Eq. (4)**) under appropriate boundary conditions. The surface tension is the extra energy cost for the formation of an interface and is defined as

$$\gamma = \frac{\left(\Omega(\rho(z)) - \Omega_{M/SS}\right)}{A} \tag{6}$$

where $\Omega_{M/SS}$ is the free energy of the coexisting melt and high density stable solid phase and $A$ is the area of the interface.

In order to study the effect of the intermediate phase on density profile and surface tension, the melt and SS phases are kept at coexistence and the stability of the intermediate phase ($\Delta\omega_{MS}$) is gradually varied. After inserting the equilibrium density profiles in **Eq. (5)**, the calculated surface tensions (using **Eq. (6)**) are shown in **Fig. 2B**. We note the strong dependence of surface tension on the extent of metastabilty of the intermediate phase. As the difference between the minimum of free energy basins between the melt and intermediate phase decreases the effect of wetting becomes more pronounced. This is reflected in both the density profile as well as surface tension. We note the significant decrease in the interfacial surface tension on increasing the stability of the intermediate phase. This decrease in surface tension is a consequence of the enhanced wetting of the high density solid interface with an intermediate density interface.

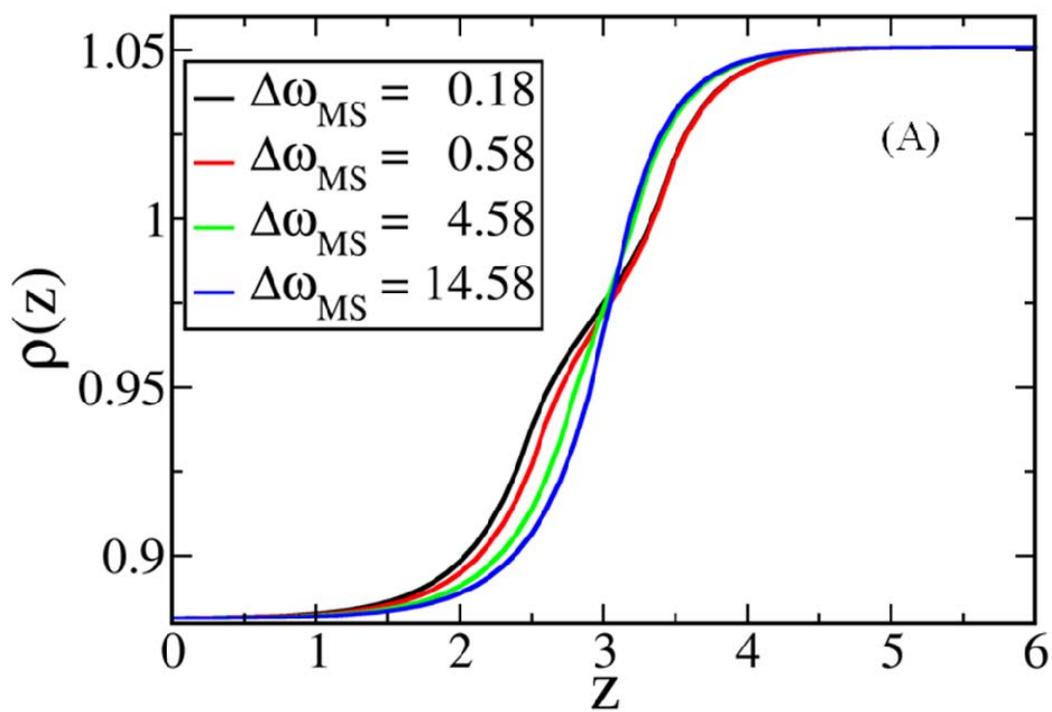

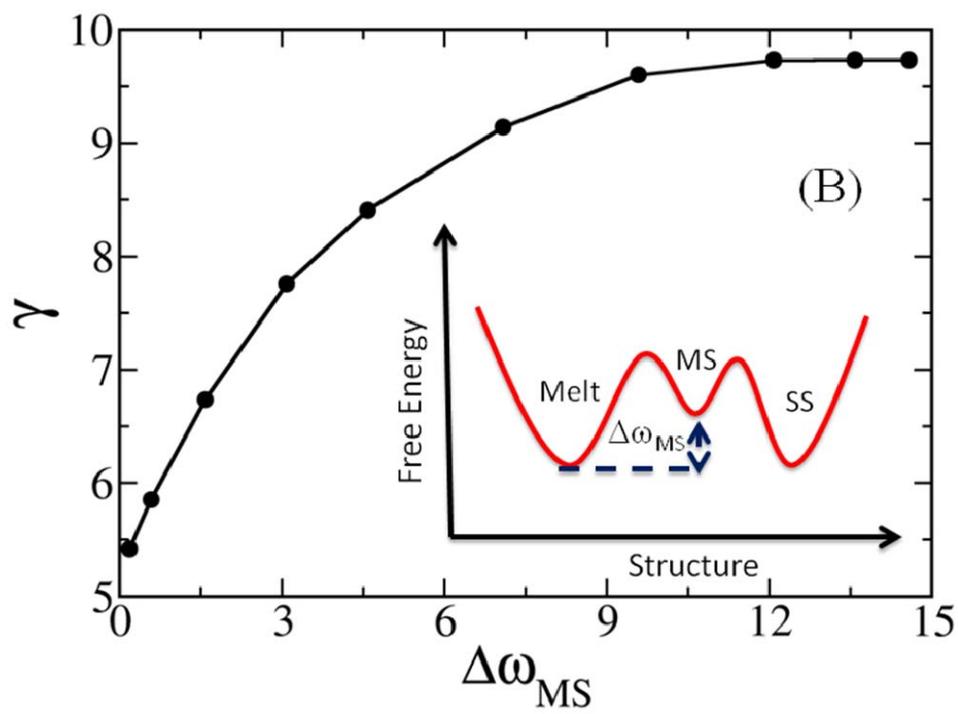

**Figure – 2.** **(A) The calculated density profiles between coexisting low density phase (melt) and the high density stable solid (SS) phase for different stabilities of metastable intermediate solid (MS) phase with respect to melt ($\Delta\omega_{MS}$, shown in inset of Fig. (B)). The free energy gap between melt and SS is fixed and stability of MS phase is gradually varied. Note the increased wetting effects on increasing stability of the intermediate phase. (B) Dependence of surface tension of melt-SS interface on extent of metastability of intermediate solid.**

### C. Crossover from wetting to Ostwald step rule

At a particular supersaturation we can evaluate the nucleation barrier as well as the density profile of the critical nucleus by solving the Euler-Lagrange equations associated with following condition,

$$\frac{\delta\Omega[\rho(r)]}{\delta\rho(r)} = 0 , \qquad (7)$$

where $\Omega[\rho(r)]$ is the grand canonical free energy functional corresponding to nucleus of daughter phase with density profile $\rho(r)$. The density profiles of critical nuclei (shown in **Fig. 3**) are obtained by solving the corresponding Euler-Lagrange equations under appropriate boundary conditions. In order to understand the effects of the intermediate phase on the composition of critical nucleus, we have fixed the free energy gap (supersaturation) between parent melt and SS phases and gradually varied the stability of the intermediate (MS) phase with respect to parent melt ($\Delta\omega_{MS}$). This construction allows us to reveal solely the effect of wetting by the intermediate phase on the composition of the critical nucleus and corresponding energy cost. In contrast, in reality this effect cannot be quantified directly since the stability of the intermediate phase is varied by changing supersaturation which also affects the stability of the SS phase. As a result, we always observe two superimposed effects on nucleation barrier – wetting by the intermediate phase and the effect due to increased stability of SS phase.

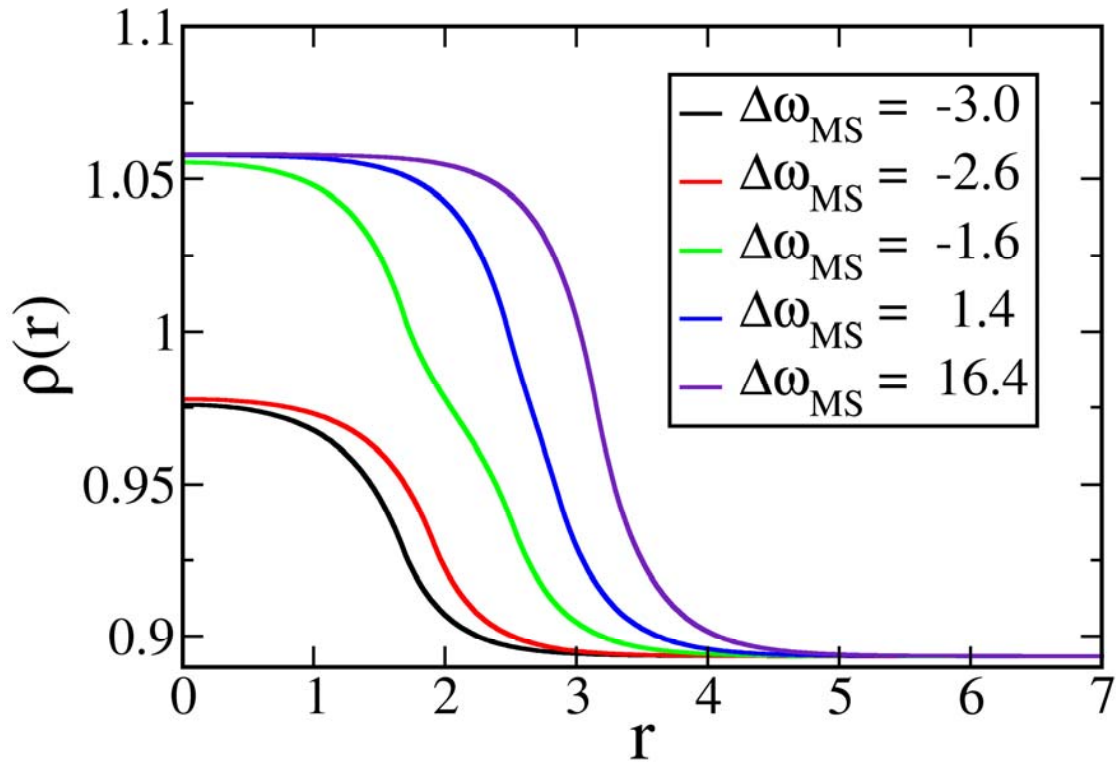

**Figure – 3. Calculated density profiles of critical nuclei for different stabilities of the intermediate phase at a fixed supersaturation (free energy difference between the melt and stable solid phases is kept fixed). Note the sudden change in density profile below a certain value of $\Delta\omega_{MS}$.**

As evident from **Fig. 3**, at a fixed supersaturation (between the low density melt and high density stable solid) density profiles for critical nucleus at different depths (extent of stability with respect to parent melt) of intermediate phase are shown. When the intermediate phase has minimal stability with respect to the low density melt phase, one step density profiles suggest the negligible role of intermediate phase in the construction of equilibrium density profile for the critical nucleus of high density phase. This indicates the absence of wetting of the nucleus of high density phase by intermediate density phase. On gradually increasing the stability of the intermediate phase we observe significant deviations in the density profile of the critical cluster of high density solid phase. These deviations indicate the change in the composition of critical cluster of SS phase by an intermediate solid phase (MS). On further increasing the stability of the intermediate phase we observe a transition where a critical cluster of intermediate phase appears inside the bulk metastable melt phase. This is the Ostwald step rule scenario, where transition from metastable phase to final stable phase occurs via many intermediate phases. *Thus on increasing the stability of the intermediate phase we observe a crossover from the wetting enhanced one step transition to the sequential two step (following Ostwald step rule) transition.*

In **Figure 4** we have shown the dependence of nucleation barrier from melt to stable solid and melt to metastable intermediate solid phases on supersaturation (pressure). At high supersaturation we note the crossover in the free energy barrier for nucleation of metastable and stable phases. At low supersaturation the nucleation free energy barrier for the stable solid phase in presence of metastable solid phase is lower than the nucleation barrier for the metastable phase. This is due to the wetting of a cluster of stable solid phase by an intermediate metastable solid phase. At low supersaturation the melt phase directly goes to the stable solid phase without encountering the bulk metastable solid phase. However, at large supersaturation, the nucleation barrier for the metastable solid phase is lower than the stable solid phase. The melt first undergoes a transition to the metastable solid phase followed by a transition from the metastable to the stable solid phase (Ostwald step rule). This crossover is also observed by Oxtoby *et al*. **[12]** in the case of one metastable solid phase, although they did not mention explicitly the Ostwald step rule type scenario. This crossover has important consequences on the pathway of phase transition in many simple and complex systems. In the next section we shall discuss the more generalized free energy landscape having multiple minima and its effect on pathways of phase transition.

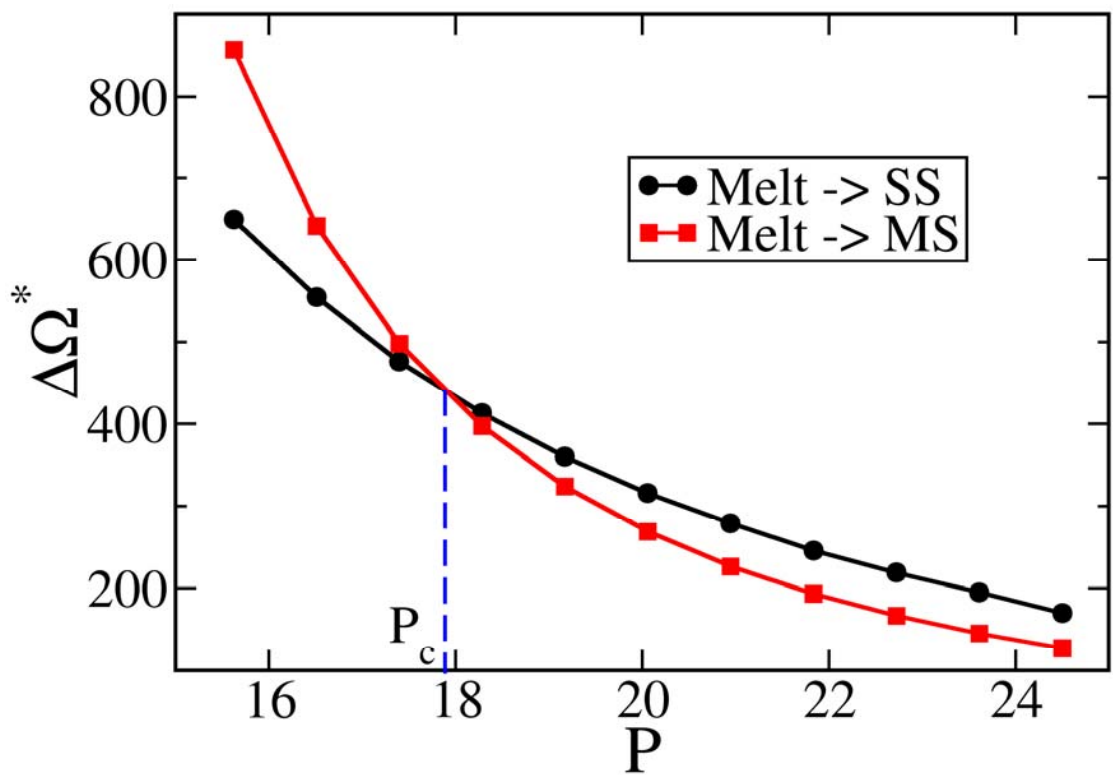

**Figure – 4. The crossover behavior. Dependence of free energy barrier of nucleation on pressure (P) is shown. The red line with filled squares indicates the free energy barrier of nucleation from melt to intermediate solid and the black line with filled circles indicates from melt to stable solid. Crossover in free energy barrier indicates transition from wetting mediated nucleation of SS to Ostwald step rule.**

### III.    Multiple metastable phases: Surface tension and nucleation barrier

## A. Surface tension

In this section we shall generalize the above discussion for the case of multiple metastable intermediate phases. For simplicity we have first considered a case of two intermediate phases. A schematic illustration of the complex free energy landscape consisting two metastable intermediate phases is provided in **Fig. 5** and **Fig. 7**. In **Fig. 5** we have shown the free energy landscape at supersaturation where melt and stable solid phase coexist along with a schematic diagram of the interface between the melt and the stable solid phase at coexistence wetted by two intermediate phases. In order to study the effects of two intermediate phases using DFT we have constructed the Helmholtz free energies of different phases (melt, M1, M2 and SS) similar to the one presented in **Eq. (2)** with following parameter values: $a_M = 1500$, $a_{M1} = 2000$, $a_{M2} = 2000$, $a_{SS} = 2500$, $\rho_M = 0.88$, $\rho_{M1} = 0.937$, $\rho_{M2} = 0.992$, $\rho_{SS} = 1.05$, $f_{M,0} = 0.0$, $f_{M1,0} = 0.08$, $f_{M2,0} = 1.0$ and $f_{SS,0} = 0.80$. In **Fig. 6**, we have plotted the equilibrium density profile for the proposed free energy surface at coexistence of melt and stable solid phases. The density profile is wetted by two metastable intermediate metastable solid phases. Similar to the earlier case (case of one intermediate phase) one would observe an enhanced wetting of interface on increasing the stability of intermediate phases. The width of the intermediate phase participating in the wetting will depend on the stability and position of the respective intermediate phase with respect to coexisting melt or SS phases.

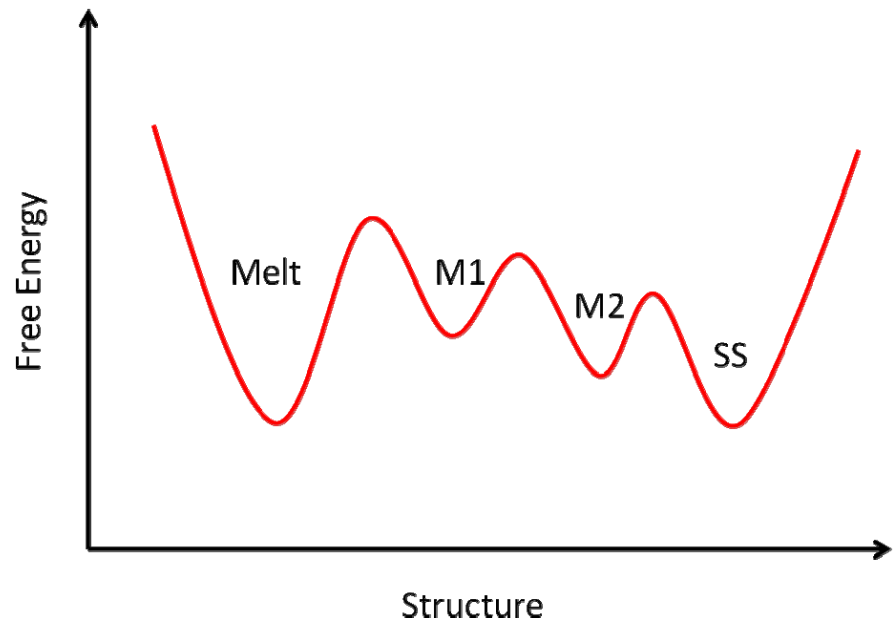

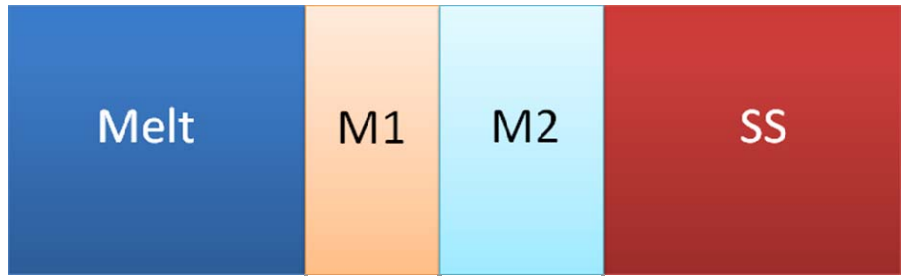

**Figure – 5.** A schematic illustration of complex energy landscape consisting two metastable intermediates is shown when the initial melt and final stable solid phases are at coexistence. A schematic illustration of the planar interface between melt and stable solid phase wetted by two intermediate phases M1 and M2 is also shown.

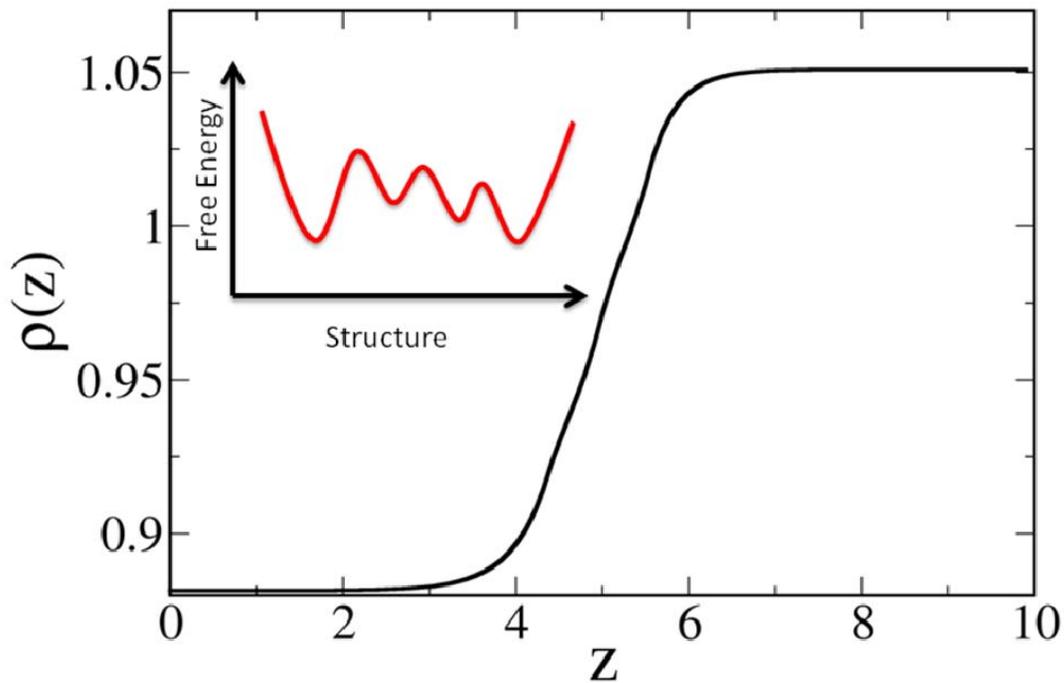

**Figure – 6. The calculated density profile for the case of two metastable intermediate phases (M1 and M2 and a schematic plot of free energy is shown in the inset) at coexistence of melt and stable solid phases.**

B. **Nucleation barrier**

In **Fig. 7**, we have shown a schematic energy landscape under supersaturated condition. Initial metastable melt is separated and stable solid phase is separated by multiple intermediate metastable (with respect to stable solid) solid phases. In the case of two metastable intermediate phases there can be three possibilities – (i) All intermediate phases participate in the wetting of the nucleus of the stable solid phase in bulk metastable melt (shown in **Fig. 7**). (ii) At first, the melt completely transforms to one of the metastable states, however, the other intermediate phase participates in wetting. This is the case of partial Ostwald step rule. (iii) System sequentially moves from melt to the stable solid phase via intermediate metastable phases. This is the case of complete Ostwald step rule. We must note that in the case of one intermediate case only two scenarios (wetting induced one step and sequential Ostwald step rule) scenarios are possible.

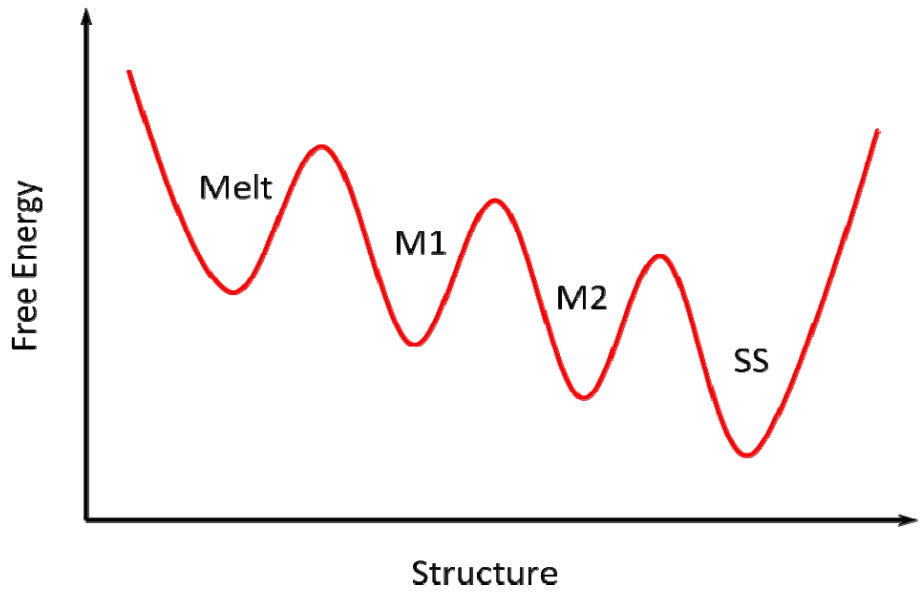

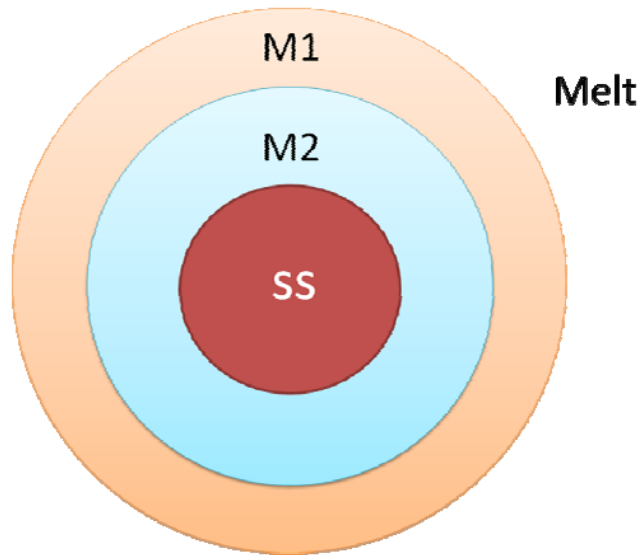

**Figure –7. A schematic representation of complex energy landscape consisting two metastable intermediates under super-saturated condition is shown. A growing nucleus wetted by intermediate phases is also shown schematically.**

In order to find the conditions under which we can realize the above mentioned different cases, we have plotted the nucleation barrier with varying supersaturation. In **Fig. 8**, we have plotted the free energy barrier for nucleation of three different phases – M1, M2 and SS from metastable melt phase. The black line indicates the supersaturation dependent free energy barrier for nucleation of stable phase in bulk metastable melt, the red line indicates the same for the nucleation of M2 and the blue line indicates the same for the nucleation of M1. We note that multiple crossovers in the free energy barriers for nucleation of different phases in melt. This has important consequence on deciding the pathways of transition and discussed in details in the next section.

At low supersaturation (pressure), the free energy barrier for nucleation of SS is lower than both of M1 and M2. That is at low supersaturation the melt phase will directly undergo a transition to the stable phase without encountering any bulk intermediate metastable (M1 and M2) phases. The role of the intermediate phases is to reduce the free energy barrier for the nucleation of stable phase by wetting. Thus not encountering a metastable intermediate phase during transition does not discard the possibility of existence of metastable states. On increasing the supersaturation we observe a crossover in the free energy barrier of stable solid and M2. Beyond the crossover point ($P_{c1}$) the system will directly undergo a transition to the M2 followed by transition to SS. The critical cluster of M2 is wetted by the M1 phase and thus reduces the free energy barrier for transition to M2. This is the partial Ostwald step rule scenario where system is exploring only one intermediate state (M2) before transferring to the final stable solid. On further increasing the supersaturation we observe another crossover where nucleation free energy barrier for M1 is lower than both M2 and SS. Beyond $P_{c3}$, the melt will first transform to M1 followed by transformation to other phases (M2 and SS). This way by modulating the free energy landscape, one can create a situation where the system will sequentially move from one metastable to another before moving to the final stable solid phase. This is the case of complete Ostwald rule.

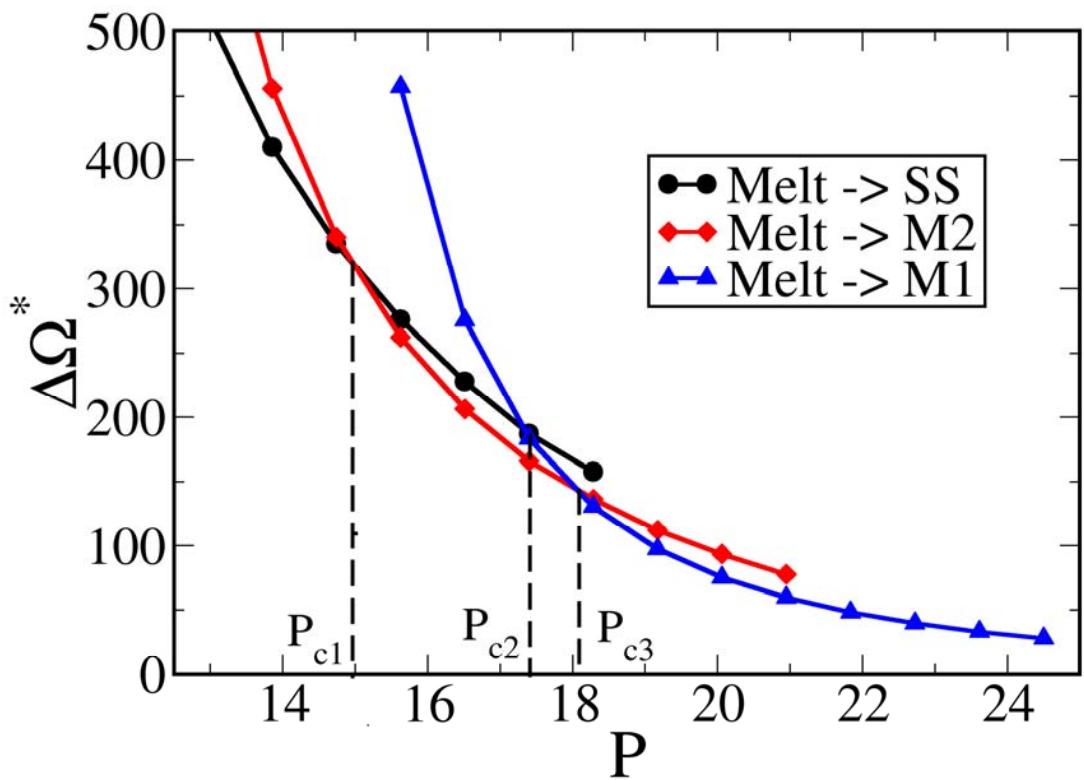

**Figure – 8. The computed free energy barriers of nucleation of different phases from melt are shown. Note the multiple crossovers at different pressure indicated by vertical dotted lines ($P_{c1}$, $P_{c2}$, $P_{c3}$).**

The different scenarios (discussed above) indicate that the crystallization pathways are richer than the predictions of Ostwald step rule and depend on many factors such as the number, well depths and positions of the free energy minima of the intermediate phases. The well depths and positions of the free energy minima can be tuned easily by changing the thermodynamic parameters such as temperature and pressure. The number of intermediate phases depends on the system under consideration. Thus change of thermodynamic parameters has profound effect on the selection of the pathways of the crystallization. Recently, Whitelam and coworker have also observed a rich crystallization pathway in the case of patchy colloidal model system **[39]**.

### C. Dependence of surface tension on multiple intermediate phases coexisting with each other

In this section we shall discuss the dependence of surface tension on the number of intermediate coexisting phases. The initial melt and final stable solid phase is separated by $N$ coexisting intermediate phases. The grand potential density difference (relative to the bulk initial phase) of melt ($\Delta\omega_M$), $i^{th}$ intermediate ($\Delta\omega_i$) and the stable solid phase ($\Delta\omega_{SS}$) is given as

$$\Delta\omega_M = \frac{1}{2}k(\rho-\rho_M)^2$$
$$\Delta\omega_i = \frac{1}{2}k(\rho-\rho_i)^2 \qquad (8)$$
$$\Delta\omega_{SS} = \frac{1}{2}k(\rho-\rho_{SS})^2$$

where for simplicity we have assumed that the curvatures of the free energy surfaces are same for all phases. $\rho_M$ and $\rho_{SS}$ are the equilibrium densities of melt and stable solid phase, respectively. $\rho_i$ is the equilibrium density of $i^{th}$ intermediate phase and is given as $\rho_M + i\Delta\rho$, where $\Delta\rho = (\rho_{SS} - \rho_M)/(N+1)$. Following Cahn-Hilliard **[40]**, the work of formation of the critical nucleus is given as

$$\Omega = \int d\mathbf{r} \left[ \Delta\omega(\rho(\mathbf{r})) + c(\nabla\rho)^2 \right] \tag{9}$$

where c is related to the correlation length. Using the analytical expression of the surface tension originally derived by Cahn and Hilliard **[40]**, $\gamma = 2\sqrt{c} \int_{\rho_i}^{\rho_f} \sqrt{\Delta\omega} d\rho$, a relation between the wetted surface tension ($\gamma_{M/SS}^w$) and without wetting ($\gamma_{M/SS}$) can be easily derived and given as

$$\gamma_{M/SS}^w = \gamma_{M/SS} \frac{2N+1}{(N+1)^2}. \tag{10}$$

where $\gamma_{M/SS} = \frac{\sqrt{2kc}}{4}(\rho_{SS} - \rho_M)^2$, here $c$ is the coefficient of the square gradient term and $k$ is the curvature of the grand potential. From the above expression it is quite evident that for larger $N$ the surface tension decreases as inverse of $N$ and as discussed earlier, this has important consequence on the nucleation and growth processes in complex and disordered systems.

### IV. Generalized free energy functional: two order parameter description

For crystallization a two order parameter (density and order) description is often necessary. For simplicity we consider the case with only one intermediate metastable solid phase (referred as MS) between the fluid (melt) and final stable solid phase (SS). We follow Oxtoby in describing the free energy functional of the inhomogeneous phase, characterized by position dependent order parameters **[38]**. The proposed free energy functionals for three phases, fluid (F), intermediate metastable solid (MS) and stable solid (SS) are

$$\Omega_i[\rho(\mathbf{r}), m(\mathbf{r})] = \int d\mathbf{r} \left[ f_i(\rho(\mathbf{r}), m(\mathbf{r})) - \mu\rho(\mathbf{r}) \right] + \frac{1}{2} \int d\mathbf{r} \left[ K_{\rho i} (\nabla\rho(\mathbf{r}))^2 \right] + \frac{1}{2} \int d\mathbf{r} \left[ K_{mi} \rho_S^2 (\nabla m(\mathbf{r}))^2 \right] \quad (11)$$

where $f_i$ is a local Helmholtz free energy density function of the average number density $\rho(\mathbf{r})$ and structural order parameter $m(\mathbf{r})$ and $\mu$ is the chemical potential. Here '$i$' indicates respective phases. The square gradient terms account for the nonlocal effects in the system due to inhomogeneity in density and structural order parameters. $K_{\rho i}$ and $K_{mi}$ are related to correlation lengths for $\rho$ and $m$. Following Talanquar and Oxtoby **[38]** the Helmholtz free energy density for homogeneous fluid is

$$f_f(\rho, m) = k_B T \rho \left[ \ln\rho - 1 - \ln(1 - b\rho) \right] - a\rho^2 + k_B T \alpha_l m^2 \quad (12)$$

The above free energy functional is a generalization of van der Waal's free energy functional for a two order parameter description. In a similar spirit one can also write the Helmholtz free energy functional for solids (metastable and stable) as

$$f_j(\rho, m) = k_B T \rho \left[ \ln\rho - 1 - \ln(1 - b\bar{\rho}) \right] - a\rho^2 + k_B T \left[ \alpha_{1j} m^2 + \alpha_{2j} \right] \quad (13)$$

where $j$ = MS for metastable intermediate solid and SS for stable solid. Here $k_B$ is the Boltzmann's constant, $T$ is the absolute temperature, $a$ and $b$ are the van der Waal's parameters and account for the effect of interactions between dissolved molecules. $\bar{\rho}$ is weighted average density and is given as

$$\bar{\rho} = \rho\left[1-\alpha_{3j}m(\alpha_{4j}-m)\right]. \tag{14}$$

The value of the parameters are, $a = 1.0$, $b = 1.0$, $\alpha_L = \alpha_{1MS} = \alpha_{1SS} = 0.25$, $\alpha_{2MS} = 1.5$, $\alpha_{3MS} = 0.22$, $\alpha_{4MS} = 1.85$, $\alpha_{2SS} = 2.0$, $\alpha_{3SS} = 0.30$, $\alpha_{4MS} = 2.0$, $K_{\rho i} = a/2$ and $K_{mi} = a/8$. A contour plot of free energy functions given by **Eq. (12)** and **Eq. (13)** are shown in **Fig. 9.**

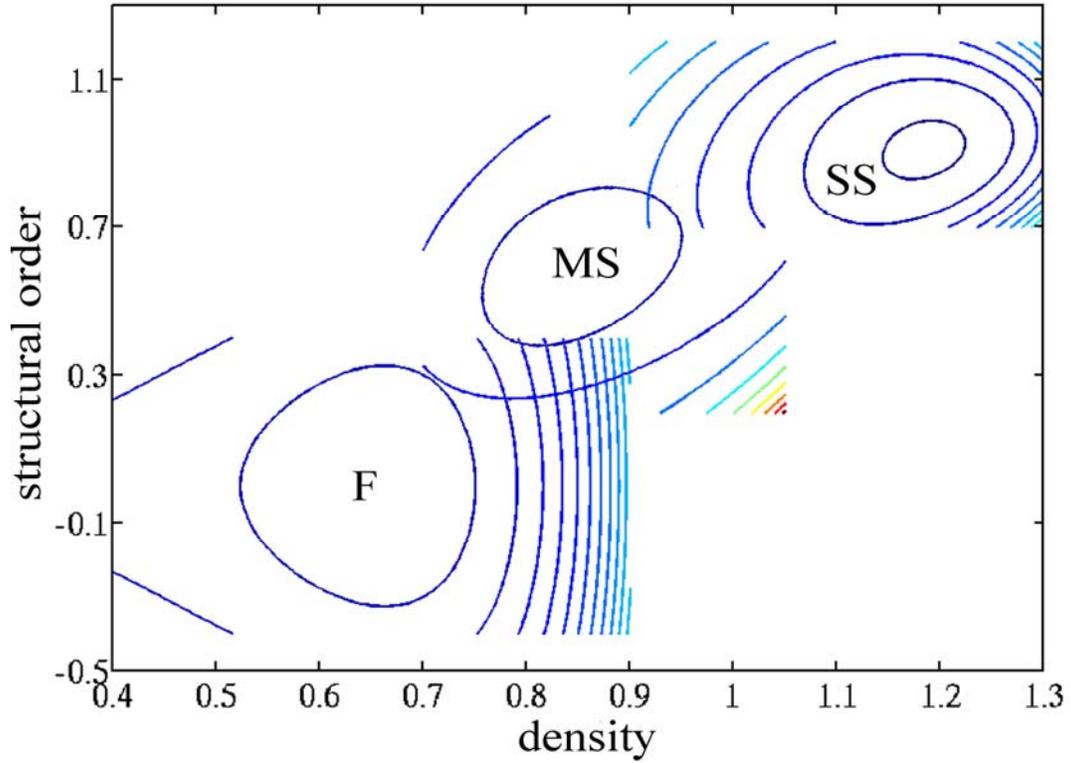

**Figure – 9. A contour diagram of two dimensional free energy surfaces given by Eq. (12) and Eq. (13) is shown. F stands for fluid phase, MS for metastable intermediate solid and SS for stable solid.**

### A. Phase diagram

The density and structural order of coexisting phases can be determined by equating chemical potential and thermodynamic grand potential density (pressure) of the two phases.

$$\mu_\alpha(\rho_\alpha) = \mu_\beta(\rho_\beta) \text{ and } \omega_\alpha(\rho_\alpha) = \omega_\beta(\rho_\beta) \tag{15}$$

where $\mu_i = \left(\dfrac{\partial f_i(\rho,m)}{\partial \rho}\right)_T$ and $\omega_i = f_i - \mu_i \rho_i$.

The above two conditions ensure that the system is in both thermodynamic and mechanical equilibrium. **Fig. 10** shows the results of quantitative calculation of the coexistence between gas-liquid, liquid - intermediate metastable solid (L-MS), liquid-stable solid (L-SS) and also between metastable and stable solid. As gas-liquid interfacial surface tension depends strongly on the order parameter difference, one can qualitatively conclude that the surface tension between the fluid and stable solid will be larger than the fluid and metastable solid phase. More accurate quantitative results for surface tension for interfaces between fluid-metastable solid and fluid-stable solid are discussed in next section.

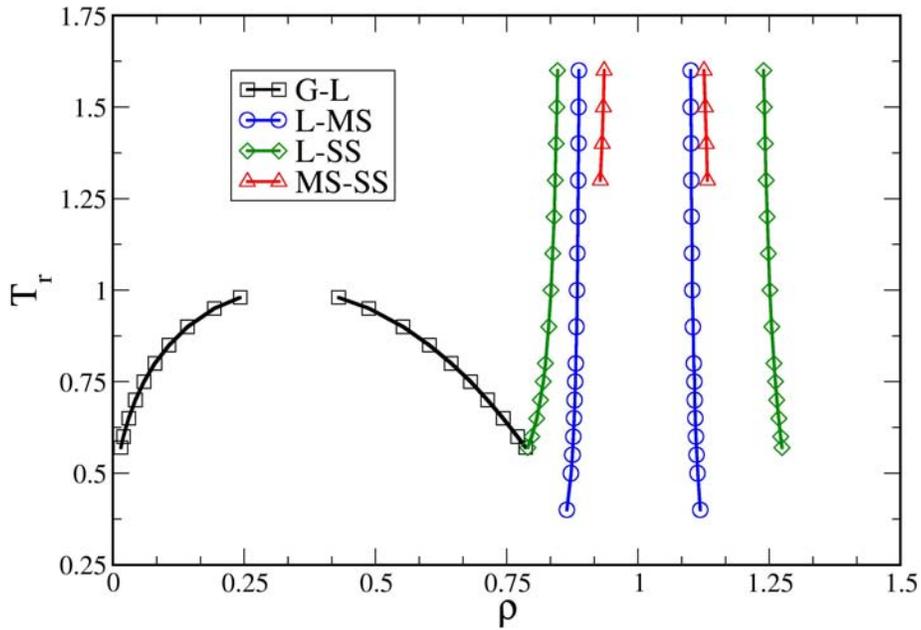

**Figure – 10.** The computed phase diagram for the proposed free energy surfaces in reduced temperature $T_r = T/T_c$. The black lines (with squares) indicate the gas-liquid coexistence. Green lines (with diamond) indicate the coexistence between fluid (F) and stable solid (SS) and blue lines (with circles) indicate the coexistence between fluid (F) and metastable solid (MS). Red lines (with triangle up) indicate the coexistence between intermediate metastable solid (MS) and stable solid (SS).

## B. Surface tension

For the phase diagram shown in **Fig. 10**, we can evaluate the values of the surface tension between the coexisting phases (fluid-stable solid (L-SS) and fluid-metastable solid (L-MS)) for a planar interface along z-axis by solving the Euler-Lagrange equations associated with following equilibrium conditions

$$\frac{\delta \Omega}{\delta \rho(z)} = 0 \quad \text{and} \quad \frac{\delta \Omega}{\delta m(z)} = 0 \ , \tag{16}$$

where $\Omega(\rho(z), m(z))$ is the grand canonical free energy functional corresponding to the inhomogeneous system with density profile $\rho(z)$ and order profile $m(z)$,

$$\Omega[\rho(z), m(z)] = \int dz \left[ f(\rho(z), m(z)) - \mu \rho(z) \right] + \frac{1}{2} \int dz \left[ K_\rho (\nabla \rho(z))^2 \right] \\ + \frac{1}{2} \int dz \left[ K_m \rho_s^2 (\nabla m(z))^2 \right]. \tag{17}$$

Here $f = \min \{ f_f, f_j \}$.

On minimizing the above free energy functional with respect to density and order profiles (or equivalently, solving the corresponding Euler-Lagrange equations under appropriate boundary conditions) we obtain the equilibrium density and order profiles. Equilibrium density and order profiles for L-MS and L-SS interfaces are shown in **Fig. 11.** The surface tension is extra free energy cost for the formation of an interface and is defined as

$$\gamma_{i/j} = \frac{\left(\Omega(\rho(z), m(z)) - \Omega_{i/j}\right)}{A} \quad (18)$$

where $\Omega_{i/j}$ is the free energy of the coexisting $i^{th}$ and $j^{th}$ phases and $A$ is the area of the interface. The calculated surface tension values (using **Eq. (17)** and **Eq. (18)**) for interfaces between liquid and metastable solid (L-MS) and liquid and stable solid (L-SS) at coexistence are $\gamma_{L/MS} = 6.8 \times 10^{-2}$ and $\gamma_{L/SS} = 14.7 \times 10^{-2}$ (in units of $a/b^{5/3}$).

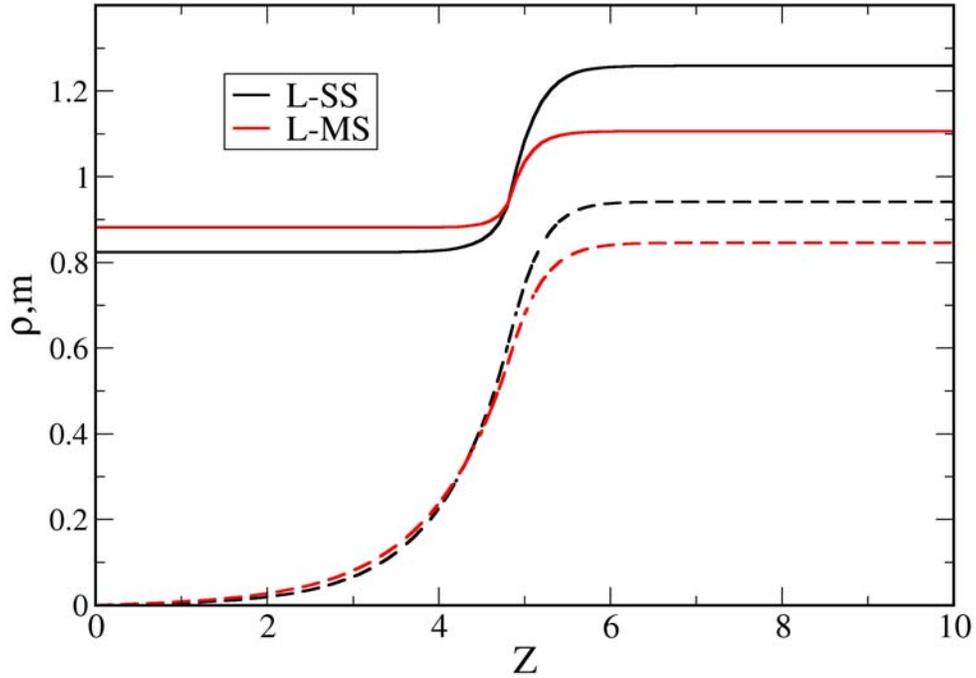

**Figure - 11. The calculated density and order profile for planar interface along z-axis at $T_r = 0.80$. Solid lines indicate the density profiles along liquid-stable solid (L-SS) and liquid-metastable solid (L-MS) interfaces, respectively. Dotted lines indicate the order profiles along liquid-stable solid (L-SS) and liquid-metastable solid (L-MS) interfaces, respectively.**

If we consider **Fig. 7** as representative of metastable crystallization, the free energy difference between different phases will follow the trend, in general, $\Delta G_{LM1} < \Delta G_{LM2} < ........ < \Delta G_{LSS}$ and the above theoretical analysis suggest, surface tension would follow $\gamma_{LM1} < \gamma_{LM2} < ........ < \gamma_{LSS}$. Physically it can be understood that the less energy is required for formation of disordered or open structured solids from the melt or sol.

## V. Kinetic scheme : vanishing polymorphs

In this section we shall briefly discuss the kinetic equations for phase transformation in presence of *one* metastable phase. Let $P_M$, $P_{MS}$ and $P_{SS}$ denote the population of metastable melt, metastable solid and stable solid phases, respectively.

$$\frac{\partial P_M(r,t)}{\partial t} = -k_{M \to MS}(r)P_M + k_{MS \to M}(r)P_{MS}$$
$$\frac{\partial P_{MS}(r,t)}{\partial t} = -k_{MS \to M}(r)P_{MS} - k_{MS \to SS}(r)P_{MS} + k_{M \to MS}(r)P_M + k_{SS \to MS}(r)P_{SS} - k_{pMS}(r)P_{MS} \quad (19)$$
$$\frac{\partial P_{SS}(r,t)}{\partial t} = -k_{SS \to MS}(r)P_{SS} + k_{MS \to SS}(r)P_{MS} - k_{pSS}(r)P_{SS}$$

where $k_{pMS}$ is the size dependent precipitation rate of metastable phase of size $r$ and $k_{pSS}$ is the precipitation rate of the stable solid phase, both from solution phase. These kinetic equations have very close resemblance with the sequential chemical reactions in presence of a sink term.

Solution of **Eq. (19)** needs values of size dependent rate constants. The size dependence of rate shall have a crossover, if we start from small, pre-critical sizes. They shall clearly depend on the degree of supersaturation. If we further assume that the rate limiting step is the nucleation step, then these transition rates can be obtained from the nucleation rate calculated here. The rates of precipitation are a bit more difficult to estimate and shall clearly depend on the size of the growing cluster. One could imagine a crossover in the size dependence here too. In a general scheme these clusters should interact with each other but neglected here. If we neglect the precipitation and the inter-cluster interaction effects, then the following scenario unfolds. When the intermediate states are metastable with respect to both the parent liquid and the final stable solid, the intermediate phases will not nucleate and shall not form, but only wet the interface. Macroscopic observation might not find any signature of the metastable phases as their only signature will be at the interface. Solution of **Eq. 19** then shall have the rates of formation of metastable phases zero, but their "hidden" effect will be in enhancing $k_{ml,ss}$ which is the rate of formation of SS from melt. In this manuscript we have called this the wetting dominated regime.

However, when the intermediate phases are of lower free energy than the parent liquid but metastable with respect to the stable solid phase, then one may be able to detect/isolate these phases. We have called this the Ostwald step rule dominated regime. But here also their appearance may be short lived if the free energy gap between them and the stable solid is large and that between them and metastable liquid is small. Thus, one shall find the appearance of vanishing polymorphs, often discussed in the literature.

## VI.   Conclusion

In Xia-Wolynes **[1]** treatment of nucleation of a liquid from a glass phase, wetting leads to a size dependent surface tension and the nucleation barrier decreases as the size of the nucleus grows. In that case, both the parent (glass) and daughter (liquid) phases are disordered and differ at thermodynamic (macroscopic) scale only by entropy. So, the precise quantification of the intermediate phases is left a bit unclear, although the results appear to be robust. The size of the critical nucleus scales as $s_c^{-2/3}$ and the activation barrier goes as $1/s_c$ (Adam-Gibbs relation). In the mosaic picture of liquid-glass transition the system is dynamically inhomogeneous with regions characterized by different entropies and dynamical properties. Configurational entropy is serving as an order parameter with certain similarity with Marcus theory where energy is order parameter.

In the present scenario of the growth of a new phase from melt, the situation is more clear. We employ an order parameter description with distinct intermediate states and we show that the surface tension inevitably decreases in the presence of metastable intermediate phases and also the free energy barrier of nucleation decreases. We find that depending on the number, depth and location of the free energy minima of intermediate phases between the parent and final phases, a great variety of situations can arise and appears to be in general conformity with the Ostwald's original hypothesis.

Unfortunately, our information about free energy surfaces of the complex systems often studied in experiments is rather poor. So in this work we have to constrain at highly general picture. We have shown that within the DFT approach the surface tension decreases as 1/N (N is the number of intermediate phases). However, the relation between the size of the critical nucleus and the surface tension remains a bit tricky. As we find that the width of the interface is sensitive to many factors.

In conclusion, we have investigated, motivated by the pioneer work of Xia and Wolynes **[1]**, influence of wetting in kinetics of 1$^{st}$ order phase transformations and provide theoretical justification of the phenomena often described by Ostwald step rule. We hope to extend the present study to evolve a more quantitative description of specific systems.

One important problem where the present ideas can find use is the much discussed crystallization of supercooled water at 231 K observed by Speedy and Angel **[41]** in 1972. While much discussion has focused on the possible existence (or absence) of a high density liquid (HDL) - low density liquid (LDL) critical point, less attention has focused on crystallization **[42, 43]**. The present work suggests that nucleation of ice can be facilitated either by wetting of the HDL-Ice interface by LDL, if LDL is indeed a metastable minimum in the order parameter space. In the alternate scenario of a critical point between HDL and LDL, the large scale fluctuations can lower the free energy barrier, as observed not-too-long ago by ten Wolde and Frenkel in case of protein crystallization **[44]**. The large scale fluctuations near a submerged critical point may be detected otherwise, as the variation of specific heat near the Widom line. From the effects on free energy barrier, it is hard to distinguish between the two scenarios as both can facilitate nucleation of ice. Only if we could study the nucleus near its critical size, we could make a distinction between the two. Nevertheless, existence of a metastable phase, like LDL water, with order intermediate between HDL and ice can greatly facilitate nucleation of ice from HDL water. Computer simulation studies of freezing of water **[42, 43, 45]** do seem to indicate ice nucleates from LDL-like regions. In fact, the presence of a metastable liquid or solid phase with intermediate order can even facilitate nucleation to such an extent that the phase transition may even look like a spinodal decomposition or the limit of stability of the liquid phase.

**ACKNOWLEDGEMENT**

It is a pleasure to dedicate this work to Professor Peter Wolynes on the celebration of his 60th birthday. From early days of his seminal work on electrochemistry, chemical kinetics, electron transfer reactions, quantum dynamics to the later stage of work on protein folding and glass transition, Peter has been a true leader in the general area of physical chemistry and chemical physics. His ideas have greatly enlightened us and benefited the field. We join many in wishing him long continued success in science, health and life. We thank Professor C. N. R. Rao and Professor J. Gopalakrishnan for helpful discussions. This work was supported in parts by grants from BRNS and DST. BB acknowledges support from JC Bose Fellowship (DST).